\documentclass[aps,prl,superscriptaddress,floatfix,twocolumn,showpacs]{revtex4-1}

\def\be {\begin{equation}}
\def\ee {\end{equation}}
\def\ber {\begin{eqnarray}}
\def\eer {\end{eqnarray}}

\usepackage{graphicx,graphics}
\usepackage{dcolumn}
\usepackage{amsmath,amssymb,amsfonts}
\usepackage{latexsym,verbatim}
\usepackage{bm}
\usepackage{color}
\usepackage[breaklinks=true,colorlinks,citecolor=blue,linkcolor=blue,urlcolor=blue]{hyperref}

\usepackage{fancybox}
\usepackage{tabularx}

\newcommand{\kB}[0]{k_\mathrm{B}}
\newcommand{\rs}[0]{r_\mathrm{s}}

\newcommand{\kF}[0]{k_\mathrm{F}}
\newcommand{\EF}[0]{E_\mathrm{F}}
\newcommand{\xc}[0]{\mathrm{xc}}

\newcommand{\cV}[0]{c_\mathrm{V}}

\begin{document}

\author{F. G. Eich}
\email[]{florian.eich@mpsd.mpg.de}
\affiliation{Max Planck Institute for the Structure and Dynamics of Matter, Luruper Chaussee 149, D-22761 Hamburg, Germany}

\author{Markus Holzmann}
\affiliation{LPMMC,  UMR 5493  of CNRS,  Universite Grenoble Alpes,   BP 166,  38042 Grenoble, France}
\affiliation{Institut  Laue  Langevin,  BP  156,  38042  Grenoble,  France}

\author{G. Vignale}
\affiliation{Department of Physics and Astronomy, University of Missouri, Columbia, Missouri 65211,~USA}

\title{Effective mass of quasiparticles from thermodynamics} 

\begin{abstract}
  We discuss the potential advantages of calculating the effective mass of quasiparticles in the interacting electron liquid from the low-temperature free energy vis-a-vis the conventional approach, in which the effective mass is obtained from approximate calculations of the self-energy, or from a quantum Monte Carlo evaluation of the energy of a variational ``quasiparticle wave function.'' While raw quantum Monte Carlo data are presently too sparse to allow for an accurate determination of the effective mass, the values estimated by this method are numerically close to the ones obtained in previous calculations using diagrammatic many-body theory. In contrast to this, a recently published parametrization of quantum Monte Carlo data for the free energy of the homogeneous electron liquid yields effective masses that considerably deviate from previous calculations and even change sign for low densities, reflecting an unphysical negative entropy. We suggest that this anomaly is related to the treatment of the exchange energy at finite temperature.
\end{abstract}


\maketitle

\section{Introduction} \label{SEC:Introduction}

According to Landau's theory of Fermi liquids~\cite{PinesBook:61}, the low-energy excitations of a homogeneous electron liquid are fermionic quasiparticles characterized by an effective mass $m^*$, a charge $-e$, and a spin $\hbar/2$. Of these three parameters only the effective mass, $m^*$, is nonuniversal, being directly affected by the electron-electron interaction.  Performing an accurate microscopic calculation of $m^*$, even in the simplest case of a homogeneous electron liquid (HEL), is a  challenge for many-body theory~\cite{GiulianiVignale:05}. The conventional approach is to perform an approximate calculation of the electronic self-energy, from which the poles of the one-electron Green's function $G(k,\omega)$ for  complex frequency $\omega$ can be found.  These poles, occurring at $\omega_k = \epsilon_k -i\gamma_k$, yield the energy ($\epsilon_k$) and the decay rate ($\gamma_k$) of quasiparticles -- the second  being much smaller than the first for quasiparticles in the vicinity of the Fermi surface $k=\kF$.  The effective mass is then computed as
\be
\frac{1}{m^*} = \frac{1}{\kF} \left. \frac{d \epsilon_k}{d k} \right\vert_{k=\kF} ~. \label{mStar}
\ee

\begin{figure}[t]
\includegraphics[width=1.0\columnwidth]{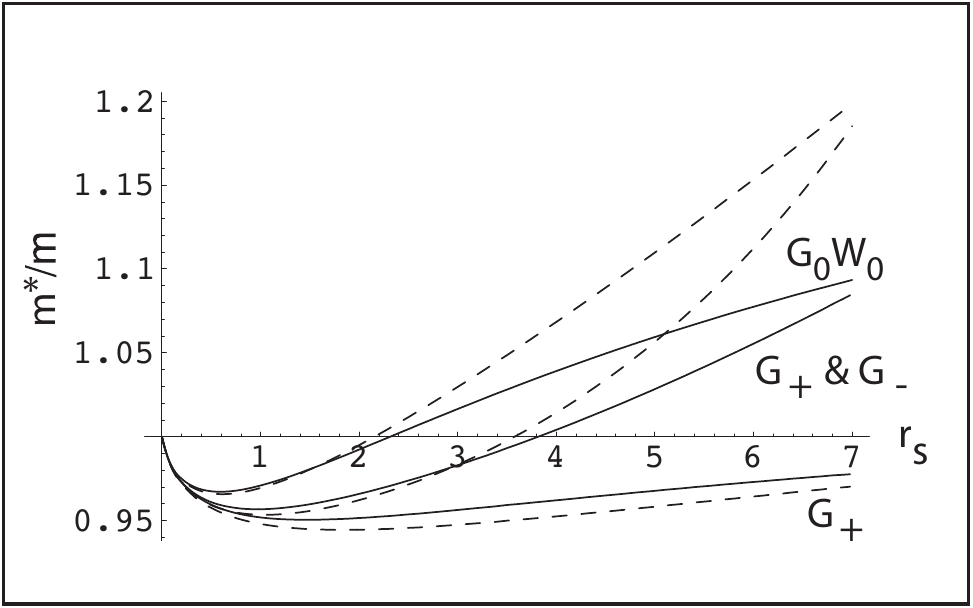}
\caption{\label{FIG:MBPT_mstar} Effective mass enhancements for the three-dimensional electron liquid.~\cite{SimionGiuliani:08} Effective masses calculated without local field factors ($G_0W_0$ approximation, also known as RPA), with local field factors $G_+$ and $G_-$ ($G_+ \& G_-$), and including only the spin-symmetric local field factor $G_+$. The solid lines represent the effective mass enhancement according to Eq.\ \eqref{mStar}, while the dashed lines correspond to the so-called ``on-shell'' approximation [cf.\ Ref.\ \onlinecite{SimionGiuliani:08} for details]. Figure adapted from Ref.\ \onlinecite{GiulianiVignale:05}.}
\end{figure}

The most popular approach  to the calculation of $m^*$ is probably the $G_0W_0$ approximation~\cite{HedinLundqvist:70}, in which the self-energy is obtained as the convolution of the noninteracting Green's function ($G_0$) with the screened electron-electron interaction ($W_0$) calculated in the random phase approximation (RPA)\cite{GiulianiVignale:05}.  Refinements of this idea include local field corrections to both the effective interaction and the screening~\cite{SimionGiuliani:08}. Figure \ref{FIG:MBPT_mstar} shows recent results that have been obtained for $m^*$ in the three-dimensional electron liquid as a function of standard parameter $\rs$ --  the average distance between electrons expressed in units of the Bohr radius $a_0$.  The slight dip of $m^*$ below the bare mass $m$ at small $\rs$ (high density) is a vestige of the dramatic suppression of the effective mass ($m^* \to 0$) that would occur if only exchange were taken into account. This happens because the Hartree-Fock self-energy has a logarithmically diverging slope at the Fermi wave vector in the zero-temperature limit. By virtue of Eq.\ \eqref{mStar}, this results in a vanishing effective mass if only exchange effects are considered.

An important message from Fig.\ \ref{FIG:MBPT_mstar} is that the plain inclusion of physical effects beyond the $G_0W_0$ approximation, is no guarantee of improvement.  For example, the inclusion of short-range correlation effects via the spin-symmetric local field factor $G_+$, which heuristically corresponds to the inclusion of vertex corrections to the self-energy, produces the curves labeled by $G_+$ in which the ratio $m^*/m$ is less than $1$ throughout the metallic density range and beyond.   However, the additional inclusion of short-range correlations via the spin-antisymmetric local field factor $G_-$, which heuristically describes the coupling of the electron quasiparticle to spin fluctuations,  yields the curve labeled ``$G_+ \& G_-$,'' which is actually close to the original $G_0W_0$ (RPA) result.     Recently, fully self-consistent $GW$ calculations for the HEL have been performed~\cite{Yan:11,VanHouckeProkofev:17}, yielding effective mass enhancements less than one for a wide range of densities ($1 \le \rs \le 30$). The results including only the spin-symmetric local field factor $G_+$, shown in Fig.\ \ref{FIG:MBPT_mstar}, exhibit a similar behavior, i.e., $m^* / m \lesssim 1$, even though they do not decrease monotonically as the self-consistent $GW$ results. By construction, $GW$ only screens the direct (Coulomb) interaction between electrons. Due to the lack of vertex corrections the effective interaction mediated by spin fluctuations (paramagnons) is missing in $GW$. Taking into account these spin fluctuations by means of the spin-antisymmetric local field factor $G_-$ enhances the effective mass, as can be seen in Fig.\ \ref{FIG:MBPT_mstar}.

Attempts to calculate the effective mass  from  quantum Monte Carlo (QMC) calculations date back to the early work of Kwon \emph{et al.} \cite{KwonMartin:94,KwonMartin:96} on the two-dimensional electron liquid.  Although QMC provides the most accurate values for the ground-state energy of the HEL~\cite{CeperleyAlder:80}, calculations of excited state properties suffer from  difficulties that are not present in the calculation of the ground state energy. One problematic point is the extrapolation to the  thermodynamic limit, since the excitations form a continuum so that energy differences at the Fermi surface are strongly affected by finite size effects \cite{HolzmannCeperley:09,HolzmannCeperley:11}. 
Furthermore, it is not clear if the  QMC methodology of Refs.\ \cite{MaezonoNeeds:03,DrummondNeeds:13a,*DrummondNeeds:13b} addressing excitation energies of small finite sized systems away from the Fermi surface can really be related to quasiparticle energies or the bandwidth. More pointedly, QMC calculations so far attempted to calculate exact energy eigenstates of the Hamiltonian for small systems, but quasiparticles are in general not exact eigenstates of the Hamiltonian -- rather they are defined as poles of the Green's function in the lower half of the complex energy plane. In particular, the finite lifetime of excitations away from the Fermi surface introduces a broadening of the quasiparticle resonance in the thermodynamic limit. Therefore, a one-to-one mapping of quasiparticle exictations, defined as the energies of the resonance peaks, to exact eigenstates of the Hamiltonian,  is not guaranteed for excitations away from the Fermi surface. Calculations of the bandwidth~\cite{MaezonoNeeds:03,DrummondNeeds:13a,*DrummondNeeds:13b} based on Landau's Fermi liquid theory may thus be problematic.

Indeed, although some QMC calculations have predicted values of the effective mass in good agreement with pre-existing GW-like theories~\cite{HolzmannCeperley:09}, no clear consensus has emerged on these values~\cite{DrummondNeeds:13a,*DrummondNeeds:13b}. 
The interpretation of experiments on simple metals such as Na -- the closest realization of the three dimensional HEL in nature \cite{HuotariOlevano:10} -- is not straightforward either due to so-called final state effects \cite{YasuharaHiguchi:99,*KuPlummer:00,*YasuharaHiguchi:00}. Electron-phonon interaction introduces further difficulties in quantitative comparisons between theory and experiment~\cite{MahanBook:90}. The measured values  of the bandwidth of sodium \cite{LyoPlummer:88}  are not in good agreement with QMC predictions~\cite{MaezonoNeeds:03}.
\begin{figure}[t]
\includegraphics[width=1.0\columnwidth]{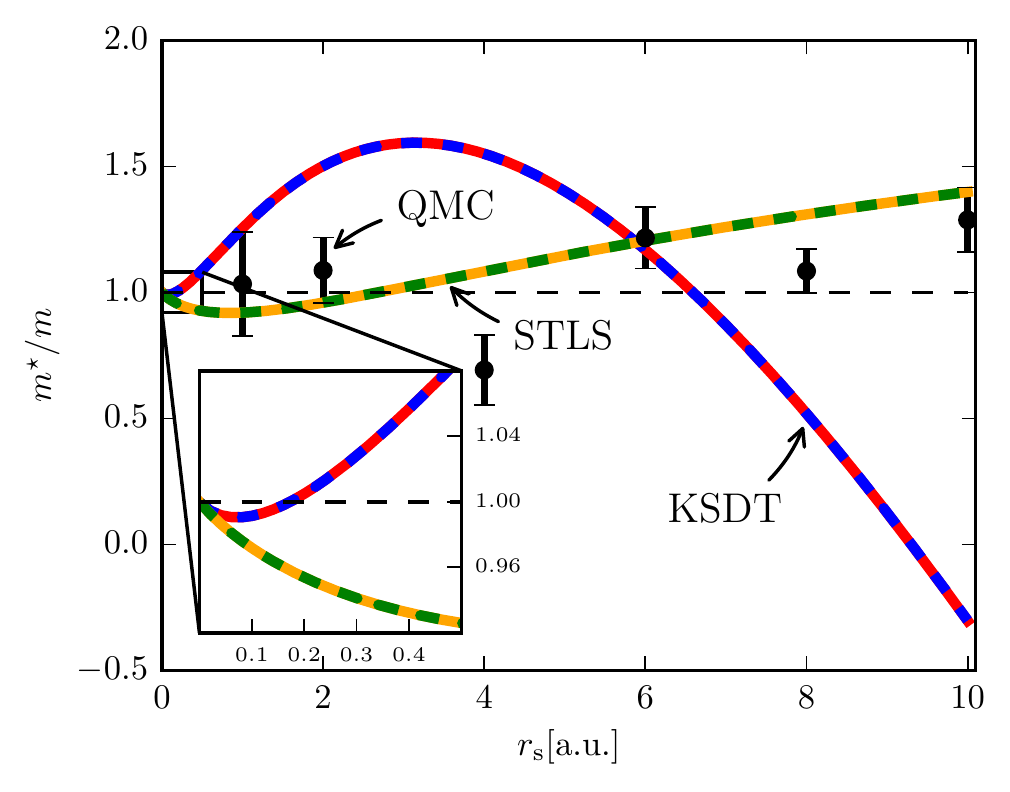}
\caption{\label{FIG:QMC_mstar} Plot depicts the effective mass enhancement computed from the raw QMC data (black dots), the KSDT parametrization of the QMC data, and the parametrization of STLS calculations at finite temperature. From the KSDT and the STLS parametrization the effective mass enhancement is obtained in two ways: 1) From the ratio of the interacting and noninteracting entropies (KSDT: red, solid line; STLS: orange, solid line). 2) From the ratio of the interacting and noninteracting heat capacity (KSDT: blue, dashed line; STLS: green, dashed line). The inset shows a zoom off the high density region ($0 \le \rs \le 0.5$). All result are obtained at $\theta = 10^{-3}$.}
\end{figure}

In this paper we propose an alternative approach, which treats ground state and excited state wave functions on equal footing and thus has the potential of being more accurate.  The method is based on the well-known property of a Fermi liquid that its  low-temperature entropy and heat capacity coincide with those of an ideal gas of Fermions of mass $m^*$.   A proof of this fact from microscopic many-body theory can be found in Ref.\ \onlinecite{AbrikosovBook} (see also Ref.\ \onlinecite{Luttinger:60}). The formal statement of this property is that the temperature dependence of the thermodynamic functions is controlled, at low temperature,  by the pole of the single-particle Green's function, i.e., the quasiparticle energy. By ``low'' temperature, we mean $\theta \equiv \kB T / \EF \ll1$ where $\EF= \hbar \kF^2 / 2 m$ is the Fermi energy determined by the Fermi wave vector, $\kF$, which, in turn is given via the density, $n$, of the electron gas, $\kF = (3 \pi^2 n)^{1/3} = (\alpha \rs a_0)^{-1}$. The numerical factor $\alpha=(4 / 9 \pi)^{1/3}$ connects the Fermi wave vector, $\kF$, with the Wigner-Seitz radius, $\rs$, in the three-dimensional electron gas. It follows  that the entropy per particle of the interacting  electron liquid, at low temperature, is given by
\be
  s = \frac{S}{N}=\frac{m^* \kF}{3\hbar^2 n}k_B^2 T ~.\label{Entropy}
\ee
$m^*$ is, strictly speaking,  the zero-temperature quasiparticle effective mass:  however, we will continue to refer to it as the effective mass, since its temperature dependence is negligible in the temperature window in which Fermi liquid theory applies. Furthermore we can use the thermodynamic identity
\be
  T \frac{\partial s}{\partial T}= \frac{\partial \epsilon}{\partial T} = \cV ~,\label{Entropy_cV}
\ee
where the derivative with respect to the temperature is taken at fixed number of particles and volume ($\epsilon$ is the internal energy per particle), to show that the entropy coincides with the heat capacity, $\cV$, \emph{in the low temperature limit}. Equation \eqref{Entropy_cV} shows that an accurate QMC calculation of the low-temperature internal energy yields the effective mass, provided that finite-size corrections can be uniformly implemented at low temperature. The exchange-correlation entropy $s_\xc$, defined as the difference between the full entropy and the noninteracting entropy $s_0=\frac{m \kF}{3\hbar^2 n}k_B^2 T $, is given by
\be
  s_\xc = \left(\frac{m^*}{m}-1\right) s_0 ~. \label{XCEntropy}
\ee
Since $m^*/m$ can be slightly less than $1$ at high density (see Fig.\ \ref{FIG:MBPT_mstar}), we can immediately conclude that $s_\xc$ (in contrast to $s$) is not necessarily positive.

The question at hand is whether more accurate calculations of $m^*$ using the method proposed above will confirm previous $G_0W_0$ calculations, or will lead to significant deviations, which would indicate the onset of strong correlation effects. It must be borne in mind in this regard that the physical basis of the $G_0W_0$ approach and its extensions is the Landau theory of Fermi liquids.  Starting from the assumption that well-defined quasiparticles exist  one arrives at $G_0W_0$  by postulating an electrostatic interaction between these quasiparticles and the collective charge density fluctuations of the liquid, and then using a standard unitary transformation to decouple the electrons from the collective degrees of freedom, generating the self-energy in the process~\cite{HamannOverhauser:66,GiulianiVignaleFL:05}. Different flavors of $GW$ differ in the form of the effective interaction between the quasiparticles and the collective degrees of freedom and yield quantitatively different results (some shown in Fig. 1), but all rely on the Fermi liquid assumption that one can draw a clear distinction between single-particle and collective degrees of freedom.  While the paradigm seems to be valid in the traditional metallic density range ($r_s < 6$) it may break down at lower density.  Hence, the importance of a calculation that does not rely on Fermi liquid assumptions.  A convincing demonstration of large qualitative  deviations from the $G_0W_0$ results for the effective mass (such as an effective mass going to infinity) would probably indicate a failure of the Fermi liquid paradigm and the need to introduce new analytical methods.

\section{Analysis and discussion} \label{SEC:Discussion}

QMC calculations of the free energy of the HEL as a function of temperature and density have been performed by Brown \emph{et al.} \cite{BrownCeperley:13a,BrownCeperley:13b,*BrownCeperley:13berratum} and improved recently \cite{SchoofBonitz:15,MaloneShepherd:16,DornheimBonitz:16}.  The first comprehensive parametrization of the QMC free energy, based on the results of Brown \emph{et al.}, has been published by Karasiev \emph{et al.} \cite{KarasievTrickey:14}, hereafter referred to as KSDT. Very recently its was shown~\cite{GrothBonitz:17} that Ichimaru's parametrization~\cite{TanakaIchimaru:86} of finite temperature STLS calculations for the uniform electron gas -- in the following simply referred to as STLS -- is a very good representation of the more recent QMC calculations presented in Refs \cite{SchoofBonitz:15,MaloneShepherd:16,DornheimBonitz:16}. The basic idea behind the STLS procedure is a self-consistent determination of the spin-symmetric local field factor $G_+$ and the static structure factor~\cite{SingwiSjoelander:68}. It has been known for some time that correlation energies obtained from the STLS structure factor are close to QMC results at zero temperature~\cite{GiulianiVignale:05}.
In the following we obtain $m^*/ m$ in three different ways: 
\begin{enumerate}
  \item from the raw QMC data of Brown \emph{et al.},
  \item using the aforementioned KSDT parametrization,
  \item employing Ichimaru's finite-temperature STLS parametrization.
\end{enumerate}

To estimate the effective mass enhancement directly from the QMC data we fit -- for each available density -- the two data points at the lowest temperature ($\theta = 0.0625, 0.125$) for the total energy to a parabola $\epsilon_\mathrm{QMC}(\theta) = C_0 + C_2 \theta^2$. Note that the lowest two temperatures presented in Ref.\ \onlinecite{DornheimBonitz:16} are $\theta = 0.5, 1.0$, which are clearly too high to extrapolate low temperature behavior. From this simple fit we obtain the heat capacity $\cV = \partial_T \epsilon(T)$ and the ratio of the QMC heat capacity and the heat capacity for the noninteracting electron gas leads to a \emph{rough estimate} for the effective mass enhancement. The black dots in Fig.~\ref{FIG:QMC_mstar} correspond to the estimates for the effective mass enhancement from the raw QMC data and the error bar is estimated from the error of the QMC total energies reported in Ref.\  \onlinecite{BrownCeperley:13a}. Clearly these estimates are too ``noisy'' for a conclusive statement but otherwise close to $1$, consistent with the $G_0W_0$ theory.  More accurate and closely spaced data for low temperatures  ($\theta < 10^{-2}$)   with carefully controlled finite size corrections will be needed before $m^*/m$ can be reliably extracted using this approach. 

Turning to the two parametrizations for the free energy, i.e., KSDT and STLS, we can extract the effective mass enhancement from the thermodynamic relation
\be\label{Entropy_FreeEnergy}
  s= -\frac{\partial f }{\partial T}~, 
\ee
where $s$ is the entropy, and $f$ the free energy per particle. Provided the temperature is low enough -- such that Eq.\ \eqref{Entropy} holds -- the mass ratio $m^*/m$ is given by the ratio of the interacting and noninteracting entropy, i.e., $m^*/m = s/s_0$. To assess whether the temperature is low enough we can alternatively compute the heat capacity (cf.\ Eq.\ \eqref{Entropy_cV}) and extract the effective mass enhancement, $m^*/m$ from the ratio of the interacting and noninteracting heat capacities. If the entropy is linear in the temperature, the ratio of the heat capacities of the interacting and the noninteracting electron gas is identical to the ratio of the interacting and noninteracting entropies.

\begin{figure}[t]
\includegraphics[width=1.0\columnwidth]{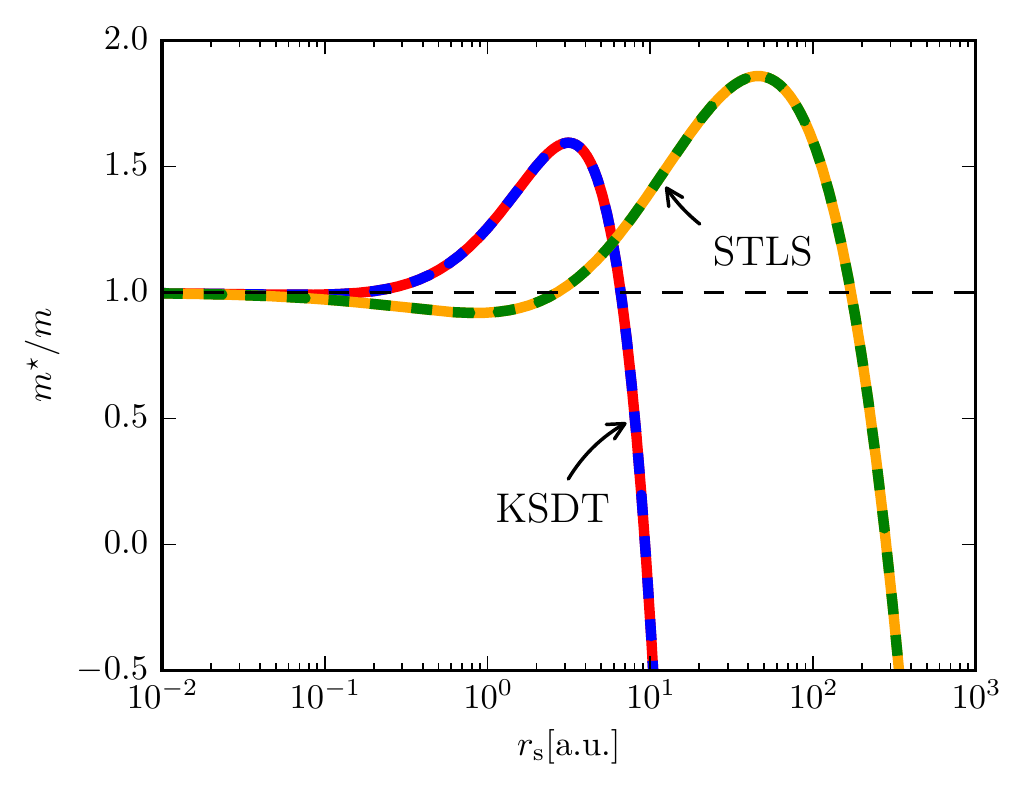}
\caption{\label{FIG:QMC_mstar_lowDensity} Similar to Fig.\ \ref{FIG:QMC_mstar} we plot the effective mass enhancement obtained from the KSDT and the STLS parametrization. Note the log-scale on the $\rs$-axis. The plot shows that both parametrizations yield an unphysical negative effective mass for low densities $\rs \gtrsim 10$ (KSDT) and very low densities $\rs \gtrsim 300$ (STLS), respectively.}
\end{figure}
 
The results, presented in Figs.\ \ref{FIG:QMC_mstar} and \ref{FIG:QMC_mstar_lowDensity}, are very surprising. The KSDT parametrization exhibits a small dip in $m^*$ at very high densities ($\rs \lesssim 1$, cf.\ inset of Fig.\ \ref{FIG:QMC_mstar}) and a large mass enhancement at somewhat lower densities ($1 \lesssim \rs \lesssim 4$).  The mass enhancement is  significantly larger than in previous calculations (compare with Fig.\ \ref{FIG:MBPT_mstar}). At even lower density we observe an odd inversion of the tendency, with the effective mass rapidly dropping below 1 and even becoming negative for $\rs \gtrsim 9.5$, which signals an unphysical negative entropy. Turning to the STLS parametrization we find mass enhancements closer to the MBPT results shown in Fig.\ \ref{FIG:MBPT_mstar}. The dip below one, due to exchange effects, extends to $\rs \sim 2.5$, which is in between the $G_0W_0$ (RPA) and the $G_- \& G_+$ results. The mass enhancement for metallic densities $\rs \lesssim 10$ predicted by the STLS parametrization seems to be quite close to the ``on-shell'' $G_0W_0$ (RPA) results (cf.\ Fig.\ \ref{FIG:MBPT_mstar}). However, for even lower densities, shown in Fig.\ \ref{FIG:QMC_mstar_lowDensity}, we see that also the STLS predicts a negative mass enhancement, indicating an unphysical negative entropy, for $\rs \gtrsim 300$. 
 
The odd behavior of the KSDT and STLS parametrizations at low temperature and low density has two potential sources: 1) inaccuracies in the underlying reference data, 2) the fitting form chosen to inter- and extrapolate the QMC (or STLS) data. Concerning the first point, we note that the QMC data has indeed an intrinsic systematic bias due to finite-size effects and the fixed-node approximation. The KSDT parametrization is based on finite-temperature QMC calculations using free particle nodes \cite{BrownCeperley:13a}, and ground-state energies \cite{SpinkDrummond:13}, which include additional backflow effects in the nodes.  Backflow nodes significantly lower energies compared to free particle nodes \cite{HolzmannEsler:03}: thus, we cannot expect that the systematic bias due to the nodal approximation cancels out. Having said this, the fact remains that the estimate for $m^* / m$ from the QMC data does not show the troubling sign reversal: thus, we believe that the fitting forms chosen for the KSDT and STLS parametrizations are responsible for producing negative entropies. In the following we discuss the probable cause of this pathology. 

\begin{figure}[t]
  \includegraphics[width=1.0\columnwidth]{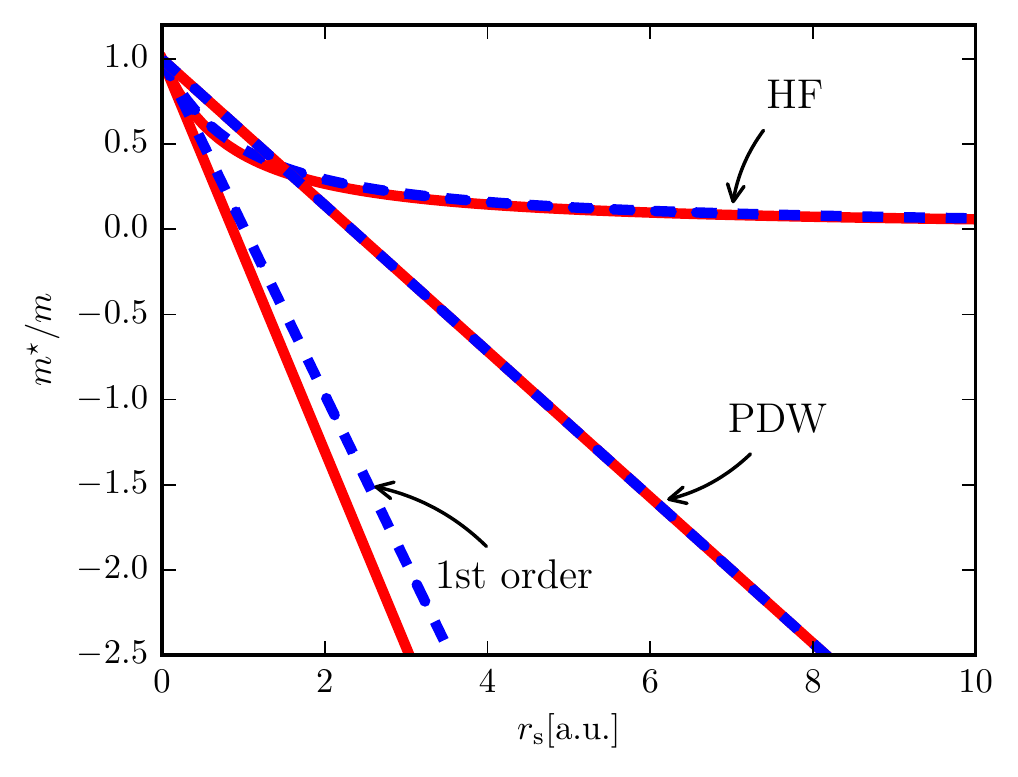}
  \caption{\label{FIG:HF_mstar} Comparison of the effective mass enhancement $m^* / m$ compute in HF and first-order exchange. The curve labeled ``1st order'' corresponds to a numerical evaluation of the first order exchange approximation, whereas the curve labeled ``PDW'' is obtained from the fit due to Perrot and Dharma-wardana. Similar to Figs.\ \ref{FIG:QMC_mstar} and \ref{FIG:QMC_mstar_lowDensity} we obtained $m^* / m$ in two ways: 1) From the ratio of the interacting and noninteracting entropies (red, solid lines). 2) From the ratio of the interacting and noninteracting heat capacity (blue, dashed lines). The notable differences between the two methods in the numerical first order and (barely visible) in the HF results are due to the non-analyticity discussed in the main text. The results are obtained at a relative temperature of $\theta=10^{-3}$.}
\end{figure}

Both KSDT and STLS incorporate as an essential component the parametrization of the first-order exchange approximation for the HEL at finite temperature proposed by Perrot and Dharma-wardana (PDW)~\cite{PerrotDharma-wardana:84}.
In Fig.\ \ref{FIG:HF_mstar}, we have plotted the mass enhancement obtained by including only the first-order exchange (computed numerically and using the PDW parametrization). It is immediately evident that the approximation fails badly, producing negative effective mass and thus an unphysical negative entropy at relatively small values of $r_s$. This unphysical behavior of the first-order exchange approximation may be the culprit for unphysical negative entropy exhibited by the KSDT parametrization and also by the STLS approximation -- albeit at considerably lower density. Surely, the coefficients in the fitting forms used in KSDT and STLS are optimized for the total exchange-correlation (free) energy, and one could argue that the unphysical first-order behavior is balanced by other terms in the functional form. However, it has to be kept in mind that the free coefficients are determined by QMC (or STLS) reference data on a restricted parameter region, which only pushes the odd behavior outside the parameter region covered by the data used in the fit.

It is  surprising, at first,  that a standard  approximation such as first-order exchange would result in an unphysical behavior of the entropy.  However, a little thought shows that this is necessarily so. While the self-consistent Hartree-Fock (HF) approximation  always produces positive entropy, the first-order exchange approximation necessarily produces negative entropy at sufficiently low density, as explained below. The difference between the two approximations is that in the HF approximation one calculates the expectation value of the interaction energy in an ensemble of single Slater determinants whose energies are self-consistently determined by the approximation itself~\cite{FetterWalecka,HongMahan:94}, whereas in the first-order exchange one calculates the expectation value of the interaction in an ensemble of Slater determinants whose energies do not include the interaction. In a uniform electron liquid, the two approximations coincide \emph{at zero temperature}, because the ground-state, with or without interactions, is given, at the mean-field level, by a single Slater determinant of plane waves  with wave vectors $k \leq \kF$. But the situation changes as soon as a non-vanishing temperature is considered, for different Slater determinants have different probabilities of occurring in the ensemble depending on whether the exchange energy  is treated \emph{self-consistently} (as in HF) or merely to first order.  Within self-consistent HF theory, the heat capacity at low temperature is approximately given by~\cite{Bardeen:36,Wigner:38}
\begin{align}
  \cV^{\mathrm{HF}} & \approx \cV^{(0)}
  \left[ 1 - \tfrac{\alpha \rs}{\pi} \log\left(\tfrac{\pi\theta}{4 \alpha \rs} \right) \right]^{-1} ~,  \label{cV_HF}
\end{align}
where $\cV^{(0)}= \kB \frac{\pi^2}{2} \theta$ is the noninteracting heat capacity and the correction factor is the effective mass enhancement of the HF approximation at low temperatures. It is evident that this expression, contains all powers of $x \propto \rs\log(\theta / \rs)$. This is because HF is a self-consistent approximation, not a perturbative expansion in $\rs$.  Notice that the HF heat capacity from this formula is positive for all $r_s$, as can be seen in Fig.\ \ref{FIG:HF_mstar}, and so is the entropy. This should be contrasted to the first-order exchange approximation, which for low temperatures is 
\begin{align}
  \cV^{(1)} & \approx \cV^{(0)} \left[ 1 + \tfrac{\alpha \rs}{\pi} \log\left(\theta\right) \right] ~.\label{cV_1}
 \end{align}
This apparently reasonable expression becomes inevitably negative for sufficiently large $r_s$ when $\theta<1$. Thus, we suggest that the negative entropy of the KSDT and STLS parametrizations arises from the inclusion of the first-order approximation to the exchange, which is necessarily unphysical at large $r_s$ and low $\theta$. In Fig.\ \ref{FIG:HF_mstar} we show the numerical results for $m^*/m$ for, both, the first-order exchange and the HF approximation. Note that the different values obtained for $m^*/m$ from the entropy ratio and the heat capacity ratio is due to the logarithmic dependence on the relative temperature  $\theta$. In addition we depict the PDW result, which is a parametrization of first-order exchange. The difference between the numerical results and the PDW parametrization is due to the fact that PDW is supposed to be an accurate parametrization at moderate temperatures, i.e., for $0.1 < \theta < 10$~\cite{PerrotDharma-wardana:84}, whereas we show results at $\theta=10^{-3}$. We have verified that PDW is much closer to the numerical first-order results for $0.1 < \theta < 10$. Furthermore there are recent parametrizations of first-order exchange, which are closer to the numerical values~\cite{KarasievTrickey:15}. The generic feature of the first-order results for the effective mass enhancement -- independent of whether they are obtained from the PDW parametrization or numerically, and irrespective of whether the entropy or the heat capacity ratio is used -- is that it is given by a straight line (as function of $\rs$) starting at $m^\star / m = 1$ with a negative slope (cf.\ Eq.\ \eqref{cV_1}). Accordingly, $m^\star / m$ is bound to become negative at low densities. We stress that the HF approximation always produces a physical $m^\star / m > 0$, which can be clearly seen in Fig.\ \ref{FIG:HF_mstar}. First-order exchange and HF only become similar at high densities ($\rs \lesssim 1$).

\section{Summary and Conclusions} \label{SEC:Summary}

We proposed to extract the effective mass for the quasiparticles in the HEL from QMC calculations at finite temperature using thermodynamic relations. However, we find that the currently available QMC data~\cite{BrownCeperley:13a,DornheimBonitz:16} are too sparse in the low temperature region for a quantitative evaluation of the effective mass. To circumvent this issue we employed a recent parametrization~\cite{KarasievTrickey:14}, which extrapolates the QMC data to arbitrary temperature and densities, to obtain the effective mass enhancement. We observe that the parametrization yields unphysical negative mass enhancements for low densities, a problem related to the fact that the current parametrization produces negative entropies, which has been pointed out earlier~\cite{BurkePribramJones:16}. Using a different parametrization of the free energy for the HEL -- based on STLS calculations at finite temperature -- we find much improved results for metallic densities. However, also the STLS parametrization exhibits unphysical, negative entropies at low densities. We trace back this oddity to the fact that both parametrizations incorporate first-order exchange as high density limit, which unfortunately spoils the low density behavior. We argue that replacing first-order exchange by fully self-consistent HF will be the cure for this pathological behavior, while still preserving the correct high density limit~\footnote{The code used to obtain thermodynamic properties of the electron gas at finite temperature within the Hartree-Fock approximation can be downloaded from \url{https://github.com/fgeich/pyFTEGhf}.}.

We hope that our analysis stimulates further QMC calculations for the HEL at low, but finite temperatures, for it holds the potential of quantifying the density region in which Landau's theory for the Fermi liquid remains valid. In closing, we point out that an accurate parametrization of the free energy of the HEL, does not only provide the local density approximation for finite temperature DFT, but is also a crucial input for the adiabatic local density approximation within nonequilibrium thermal DFT, a generalization of time-dependent DFT, which allows for the \emph{ab initio} calculation of thermoelectric transport properties~\cite{EichVignale:14a,EichVignale:17}. We emphasize that in the context of nonequilibrium thermal DFT, a physically sensible low-temperature behavior of the entropy is of utmost importance. This is in contrast to finite-temperature DFT, nowadays sometimes also referred to as equilibrium thermal DFT, which focuses on the warm-dense matter regime, i.e., a regime where an unphysical low temperature behavior might be less problematic.

\emph{Acknowledgments} -- F. G. E. and G. V. acknowledge useful discussions with Sam Trickey and Valentin Karasiev. F. G. E. has received funding from the European Union's Framework Programme for Research and Innovation Horizon 2020 (2014-2020) under the Marie Sk{\l}odowska-Curie Grant Agreement No. 701796. G. V. acknowledges support from DOE Grant DE-FG02-05ER46203. 

\bibliography{effectiveMass}

%merlin.mbs apsrev4-1.bst 2010-07-25 4.21a (PWD, AO, DPC) hacked
%Control: key (0)
%Control: author (8) initials jnrlst
%Control: editor formatted (1) identically to author
%Control: production of article title (-1) disabled
%Control: page (0) single
%Control: year (1) truncated
%Control: production of eprint (0) enabled
\begin{thebibliography}{46}%
\makeatletter
\providecommand \@ifxundefined [1]{%
 \@ifx{#1\undefined}
}%
\providecommand \@ifnum [1]{%
 \ifnum #1\expandafter \@firstoftwo
 \else \expandafter \@secondoftwo
 \fi
}%
\providecommand \@ifx [1]{%
 \ifx #1\expandafter \@firstoftwo
 \else \expandafter \@secondoftwo
 \fi
}%
\providecommand \natexlab [1]{#1}%
\providecommand \enquote  [1]{``#1''}%
\providecommand \bibnamefont  [1]{#1}%
\providecommand \bibfnamefont [1]{#1}%
\providecommand \citenamefont [1]{#1}%
\providecommand \href@noop [0]{\@secondoftwo}%
\providecommand \href [0]{\begingroup \@sanitize@url \@href}%
\providecommand \@href[1]{\@@startlink{#1}\@@href}%
\providecommand \@@href[1]{\endgroup#1\@@endlink}%
\providecommand \@sanitize@url [0]{\catcode `\\12\catcode `\$12\catcode
  `\&12\catcode `\#12\catcode `\^12\catcode `\_12\catcode `\%12\relax}%
\providecommand \@@startlink[1]{}%
\providecommand \@@endlink[0]{}%
\providecommand \url  [0]{\begingroup\@sanitize@url \@url }%
\providecommand \@url [1]{\endgroup\@href {#1}{\urlprefix }}%
\providecommand \urlprefix  [0]{URL }%
\providecommand \Eprint [0]{\href }%
\providecommand \doibase [0]{http://dx.doi.org/}%
\providecommand \selectlanguage [0]{\@gobble}%
\providecommand \bibinfo  [0]{\@secondoftwo}%
\providecommand \bibfield  [0]{\@secondoftwo}%
\providecommand \translation [1]{[#1]}%
\providecommand \BibitemOpen [0]{}%
\providecommand \bibitemStop [0]{}%
\providecommand \bibitemNoStop [0]{.\EOS\space}%
\providecommand \EOS [0]{\spacefactor3000\relax}%
\providecommand \BibitemShut  [1]{\csname bibitem#1\endcsname}%
\let\auto@bib@innerbib\@empty
%</preamble>
\bibitem [{\citenamefont {Pines}(1961)}]{PinesBook:61}%
  \BibitemOpen
  \bibfield  {author} {\bibinfo {author} {\bibfnamefont {D.}~\bibnamefont
  {Pines}},\ }\href@noop {} {\emph {\bibinfo {title} {The many-body problem: a
  lecture note and reprint volume}}},\ Frontiers in physics\ (\bibinfo
  {publisher} {W.A. Benjamin},\ \bibinfo {year} {1961})\BibitemShut {NoStop}%
\bibitem [{\citenamefont {Giuliani}\ and\ \citenamefont
  {Vignale}(2005{\natexlab{a}})}]{GiulianiVignale:05}%
  \BibitemOpen
  \bibfield  {author} {\bibinfo {author} {\bibfnamefont {G.~F.}\ \bibnamefont
  {Giuliani}}\ and\ \bibinfo {author} {\bibfnamefont {G.}~\bibnamefont
  {Vignale}},\ }\href@noop {} {\emph {\bibinfo {title} {{Q}uantum {T}heory of
  the {E}lectron {L}iquid}}}\ (\bibinfo  {publisher} {Cambridge University
  Press},\ \bibinfo {address} {Cambridge},\ \bibinfo {year} {2005})\BibitemShut
  {NoStop}%
\bibitem [{\citenamefont {Simion}\ and\ \citenamefont
  {Giuliani}(2008)}]{SimionGiuliani:08}%
  \BibitemOpen
  \bibfield  {author} {\bibinfo {author} {\bibfnamefont {G.~E.}\ \bibnamefont
  {Simion}}\ and\ \bibinfo {author} {\bibfnamefont {G.~F.}\ \bibnamefont
  {Giuliani}},\ }\href {\doibase 10.1103/PhysRevB.77.035131} {\bibfield
  {journal} {\bibinfo  {journal} {Phys. Rev. B}\ }\textbf {\bibinfo {volume}
  {77}},\ \bibinfo {pages} {035131} (\bibinfo {year} {2008})}\BibitemShut
  {NoStop}%
\bibitem [{\citenamefont {Hedin}\ and\ \citenamefont
  {Lundqvist}(1970)}]{HedinLundqvist:70}%
  \BibitemOpen
  \bibfield  {author} {\bibinfo {author} {\bibfnamefont {L.}~\bibnamefont
  {Hedin}}\ and\ \bibinfo {author} {\bibfnamefont {S.}~\bibnamefont
  {Lundqvist}},\ }\href
  {http://www.sciencedirect.com/science/article/pii/S0081194708606153}
  {\bibfield  {journal} {\bibinfo  {journal} {Solid State Physics}\ }\textbf
  {\bibinfo {volume} {23}},\ \bibinfo {pages} {1 } (\bibinfo {year}
  {1970})}\BibitemShut {NoStop}%
\bibitem [{\citenamefont {Yan}(2011)}]{Yan:11}%
  \BibitemOpen
  \bibfield  {author} {\bibinfo {author} {\bibfnamefont {X.-Z.}\ \bibnamefont
  {Yan}},\ }\href {\doibase 10.1103/PhysRevE.84.016706} {\bibfield  {journal}
  {\bibinfo  {journal} {Phys. Rev. E}\ }\textbf {\bibinfo {volume} {84}},\
  \bibinfo {pages} {016706} (\bibinfo {year} {2011})}\BibitemShut {NoStop}%
\bibitem [{\citenamefont {Van~Houcke}\ \emph {et~al.}(2017)\citenamefont
  {Van~Houcke}, \citenamefont {Tupitsyn}, \citenamefont {Mishchenko},\ and\
  \citenamefont {Prokof'ev}}]{VanHouckeProkofev:17}%
  \BibitemOpen
  \bibfield  {author} {\bibinfo {author} {\bibfnamefont {K.}~\bibnamefont
  {Van~Houcke}}, \bibinfo {author} {\bibfnamefont {I.~S.}\ \bibnamefont
  {Tupitsyn}}, \bibinfo {author} {\bibfnamefont {A.~S.}\ \bibnamefont
  {Mishchenko}}, \ and\ \bibinfo {author} {\bibfnamefont {N.~V.}\ \bibnamefont
  {Prokof'ev}},\ }\href {\doibase 10.1103/PhysRevB.95.195131} {\bibfield
  {journal} {\bibinfo  {journal} {Phys. Rev. B}\ }\textbf {\bibinfo {volume}
  {95}},\ \bibinfo {pages} {195131} (\bibinfo {year} {2017})}\BibitemShut
  {NoStop}%
\bibitem [{\citenamefont {Kwon}\ \emph {et~al.}(1994)\citenamefont {Kwon},
  \citenamefont {Ceperley},\ and\ \citenamefont {Martin}}]{KwonMartin:94}%
  \BibitemOpen
  \bibfield  {author} {\bibinfo {author} {\bibfnamefont {Y.}~\bibnamefont
  {Kwon}}, \bibinfo {author} {\bibfnamefont {D.~M.}\ \bibnamefont {Ceperley}},
  \ and\ \bibinfo {author} {\bibfnamefont {R.~M.}\ \bibnamefont {Martin}},\
  }\href {\doibase 10.1103/PhysRevB.50.1684} {\bibfield  {journal} {\bibinfo
  {journal} {Phys. Rev. B}\ }\textbf {\bibinfo {volume} {50}},\ \bibinfo
  {pages} {1684} (\bibinfo {year} {1994})}\BibitemShut {NoStop}%
\bibitem [{\citenamefont {Kwon}\ \emph {et~al.}(1996)\citenamefont {Kwon},
  \citenamefont {Ceperley},\ and\ \citenamefont {Martin}}]{KwonMartin:96}%
  \BibitemOpen
  \bibfield  {author} {\bibinfo {author} {\bibfnamefont {Y.}~\bibnamefont
  {Kwon}}, \bibinfo {author} {\bibfnamefont {D.~M.}\ \bibnamefont {Ceperley}},
  \ and\ \bibinfo {author} {\bibfnamefont {R.~M.}\ \bibnamefont {Martin}},\
  }\href {\doibase 10.1103/PhysRevB.53.7376} {\bibfield  {journal} {\bibinfo
  {journal} {Phys. Rev. B}\ }\textbf {\bibinfo {volume} {53}},\ \bibinfo
  {pages} {7376} (\bibinfo {year} {1996})}\BibitemShut {NoStop}%
\bibitem [{\citenamefont {Ceperley}\ and\ \citenamefont
  {Alder}(1980)}]{CeperleyAlder:80}%
  \BibitemOpen
  \bibfield  {author} {\bibinfo {author} {\bibfnamefont {D.~M.}\ \bibnamefont
  {Ceperley}}\ and\ \bibinfo {author} {\bibfnamefont {B.~J.}\ \bibnamefont
  {Alder}},\ }\href@noop {} {\bibfield  {journal} {\bibinfo  {journal} {Phys.
  Rev. Lett.}\ }\textbf {\bibinfo {volume} {45}},\ \bibinfo {pages} {566}
  (\bibinfo {year} {1980})}\BibitemShut {NoStop}%
\bibitem [{\citenamefont {Holzmann}\ \emph {et~al.}(2009)\citenamefont
  {Holzmann}, \citenamefont {Bernu}, \citenamefont {Olevano}, \citenamefont
  {Martin},\ and\ \citenamefont {Ceperley}}]{HolzmannCeperley:09}%
  \BibitemOpen
  \bibfield  {author} {\bibinfo {author} {\bibfnamefont {M.}~\bibnamefont
  {Holzmann}}, \bibinfo {author} {\bibfnamefont {B.}~\bibnamefont {Bernu}},
  \bibinfo {author} {\bibfnamefont {V.}~\bibnamefont {Olevano}}, \bibinfo
  {author} {\bibfnamefont {R.~M.}\ \bibnamefont {Martin}}, \ and\ \bibinfo
  {author} {\bibfnamefont {D.~M.}\ \bibnamefont {Ceperley}},\ }\href {\doibase
  10.1103/PhysRevB.79.041308} {\bibfield  {journal} {\bibinfo  {journal} {Phys.
  Rev. B}\ }\textbf {\bibinfo {volume} {79}},\ \bibinfo {pages} {041308}
  (\bibinfo {year} {2009})}\BibitemShut {NoStop}%
\bibitem [{\citenamefont {Holzmann}\ \emph {et~al.}(2011)\citenamefont
  {Holzmann}, \citenamefont {Bernu},\ and\ \citenamefont
  {Ceperley}}]{HolzmannCeperley:11}%
  \BibitemOpen
  \bibfield  {author} {\bibinfo {author} {\bibfnamefont {M.}~\bibnamefont
  {Holzmann}}, \bibinfo {author} {\bibfnamefont {B.}~\bibnamefont {Bernu}}, \
  and\ \bibinfo {author} {\bibfnamefont {D.~M.}\ \bibnamefont {Ceperley}},\
  }\href {http://stacks.iop.org/1742-6596/321/i=1/a=012020} {\bibfield
  {journal} {\bibinfo  {journal} {Journal of Physics: Conference Series}\
  }\textbf {\bibinfo {volume} {321}},\ \bibinfo {pages} {012020} (\bibinfo
  {year} {2011})}\BibitemShut {NoStop}%
\bibitem [{\citenamefont {Maezono}\ \emph {et~al.}(2003)\citenamefont
  {Maezono}, \citenamefont {Towler}, \citenamefont {Lee},\ and\ \citenamefont
  {Needs}}]{MaezonoNeeds:03}%
  \BibitemOpen
  \bibfield  {author} {\bibinfo {author} {\bibfnamefont {R.}~\bibnamefont
  {Maezono}}, \bibinfo {author} {\bibfnamefont {M.~D.}\ \bibnamefont {Towler}},
  \bibinfo {author} {\bibfnamefont {Y.}~\bibnamefont {Lee}}, \ and\ \bibinfo
  {author} {\bibfnamefont {R.~J.}\ \bibnamefont {Needs}},\ }\href {\doibase
  10.1103/PhysRevB.68.165103} {\bibfield  {journal} {\bibinfo  {journal} {Phys.
  Rev. B}\ }\textbf {\bibinfo {volume} {68}},\ \bibinfo {pages} {165103}
  (\bibinfo {year} {2003})}\BibitemShut {NoStop}%
\bibitem [{\citenamefont {Drummond}\ and\ \citenamefont
  {Needs}(2013{\natexlab{a}})}]{DrummondNeeds:13a}%
  \BibitemOpen
  \bibfield  {author} {\bibinfo {author} {\bibfnamefont {N.~D.}\ \bibnamefont
  {Drummond}}\ and\ \bibinfo {author} {\bibfnamefont {R.~J.}\ \bibnamefont
  {Needs}},\ }\href {\doibase 10.1103/PhysRevB.87.045131} {\bibfield  {journal}
  {\bibinfo  {journal} {Phys. Rev. B}\ }\textbf {\bibinfo {volume} {87}},\
  \bibinfo {pages} {045131} (\bibinfo {year} {2013}{\natexlab{a}})}\BibitemShut
  {NoStop}%
\bibitem [{\citenamefont {Drummond}\ and\ \citenamefont
  {Needs}(2013{\natexlab{b}})}]{DrummondNeeds:13b}%
  \BibitemOpen
  \bibfield  {author} {\bibinfo {author} {\bibfnamefont {N.~D.}\ \bibnamefont
  {Drummond}}\ and\ \bibinfo {author} {\bibfnamefont {R.~J.}\ \bibnamefont
  {Needs}},\ }\href {\doibase 10.1103/PhysRevB.88.035133} {\bibfield  {journal}
  {\bibinfo  {journal} {Phys. Rev. B}\ }\textbf {\bibinfo {volume} {88}},\
  \bibinfo {pages} {035133} (\bibinfo {year} {2013}{\natexlab{b}})}\BibitemShut
  {NoStop}%
\bibitem [{\citenamefont {Huotari}\ \emph {et~al.}(2010)\citenamefont
  {Huotari}, \citenamefont {Soininen}, \citenamefont {Pylkk\"anen},
  \citenamefont {H\"am\"al\"ainen}, \citenamefont {Issolah}, \citenamefont
  {Titov}, \citenamefont {McMinis}, \citenamefont {Kim}, \citenamefont {Esler},
  \citenamefont {Ceperley}, \citenamefont {Holzmann},\ and\ \citenamefont
  {Olevano}}]{HuotariOlevano:10}%
  \BibitemOpen
  \bibfield  {author} {\bibinfo {author} {\bibfnamefont {S.}~\bibnamefont
  {Huotari}}, \bibinfo {author} {\bibfnamefont {J.~A.}\ \bibnamefont
  {Soininen}}, \bibinfo {author} {\bibfnamefont {T.}~\bibnamefont
  {Pylkk\"anen}}, \bibinfo {author} {\bibfnamefont {K.}~\bibnamefont
  {H\"am\"al\"ainen}}, \bibinfo {author} {\bibfnamefont {A.}~\bibnamefont
  {Issolah}}, \bibinfo {author} {\bibfnamefont {A.}~\bibnamefont {Titov}},
  \bibinfo {author} {\bibfnamefont {J.}~\bibnamefont {McMinis}}, \bibinfo
  {author} {\bibfnamefont {J.}~\bibnamefont {Kim}}, \bibinfo {author}
  {\bibfnamefont {K.}~\bibnamefont {Esler}}, \bibinfo {author} {\bibfnamefont
  {D.~M.}\ \bibnamefont {Ceperley}}, \bibinfo {author} {\bibfnamefont
  {M.}~\bibnamefont {Holzmann}}, \ and\ \bibinfo {author} {\bibfnamefont
  {V.}~\bibnamefont {Olevano}},\ }\href {\doibase
  10.1103/PhysRevLett.105.086403} {\bibfield  {journal} {\bibinfo  {journal}
  {Phys. Rev. Lett.}\ }\textbf {\bibinfo {volume} {105}},\ \bibinfo {pages}
  {086403} (\bibinfo {year} {2010})}\BibitemShut {NoStop}%
\bibitem [{\citenamefont {Yasuhara}\ \emph {et~al.}(1999)\citenamefont
  {Yasuhara}, \citenamefont {Yoshinaga},\ and\ \citenamefont
  {Higuchi}}]{YasuharaHiguchi:99}%
  \BibitemOpen
  \bibfield  {author} {\bibinfo {author} {\bibfnamefont {H.}~\bibnamefont
  {Yasuhara}}, \bibinfo {author} {\bibfnamefont {S.}~\bibnamefont {Yoshinaga}},
  \ and\ \bibinfo {author} {\bibfnamefont {M.}~\bibnamefont {Higuchi}},\ }\href
  {\doibase 10.1103/PhysRevLett.83.3250} {\bibfield  {journal} {\bibinfo
  {journal} {Phys. Rev. Lett.}\ }\textbf {\bibinfo {volume} {83}},\ \bibinfo
  {pages} {3250} (\bibinfo {year} {1999})}\BibitemShut {NoStop}%
\bibitem [{\citenamefont {Ku}\ \emph {et~al.}(2000)\citenamefont {Ku},
  \citenamefont {Eguiluz},\ and\ \citenamefont {Plummer}}]{KuPlummer:00}%
  \BibitemOpen
  \bibfield  {author} {\bibinfo {author} {\bibfnamefont {W.}~\bibnamefont
  {Ku}}, \bibinfo {author} {\bibfnamefont {A.~G.}\ \bibnamefont {Eguiluz}}, \
  and\ \bibinfo {author} {\bibfnamefont {E.~W.}\ \bibnamefont {Plummer}},\
  }\href {\doibase 10.1103/PhysRevLett.85.2410} {\bibfield  {journal} {\bibinfo
   {journal} {Phys. Rev. Lett.}\ }\textbf {\bibinfo {volume} {85}},\ \bibinfo
  {pages} {2410} (\bibinfo {year} {2000})}\BibitemShut {NoStop}%
\bibitem [{\citenamefont {Yasuhara}\ \emph {et~al.}(2000)\citenamefont
  {Yasuhara}, \citenamefont {Yoshinaga},\ and\ \citenamefont
  {Higuchi}}]{YasuharaHiguchi:00}%
  \BibitemOpen
  \bibfield  {author} {\bibinfo {author} {\bibfnamefont {H.}~\bibnamefont
  {Yasuhara}}, \bibinfo {author} {\bibfnamefont {S.}~\bibnamefont {Yoshinaga}},
  \ and\ \bibinfo {author} {\bibfnamefont {M.}~\bibnamefont {Higuchi}},\ }\href
  {\doibase 10.1103/PhysRevLett.85.2411} {\bibfield  {journal} {\bibinfo
  {journal} {Phys. Rev. Lett.}\ }\textbf {\bibinfo {volume} {85}},\ \bibinfo
  {pages} {2411} (\bibinfo {year} {2000})}\BibitemShut {NoStop}%
\bibitem [{\citenamefont {Mahan}(1990)}]{MahanBook:90}%
  \BibitemOpen
  \bibfield  {author} {\bibinfo {author} {\bibfnamefont {G.}~\bibnamefont
  {Mahan}},\ }\href@noop {} {\emph {\bibinfo {title} {Many-Particle
  Physics}}},\ Physics of Solids and Liquids\ (\bibinfo  {publisher} {Springer
  US},\ \bibinfo {year} {1990})\BibitemShut {NoStop}%
\bibitem [{\citenamefont {Lyo}\ and\ \citenamefont
  {Plummer}(1988)}]{LyoPlummer:88}%
  \BibitemOpen
  \bibfield  {author} {\bibinfo {author} {\bibfnamefont {I.-W.}\ \bibnamefont
  {Lyo}}\ and\ \bibinfo {author} {\bibfnamefont {E.~W.}\ \bibnamefont
  {Plummer}},\ }\href {\doibase 10.1103/PhysRevLett.60.1558} {\bibfield
  {journal} {\bibinfo  {journal} {Phys. Rev. Lett.}\ }\textbf {\bibinfo
  {volume} {60}},\ \bibinfo {pages} {1558} (\bibinfo {year}
  {1988})}\BibitemShut {NoStop}%
\bibitem [{\citenamefont {Abrikosov}\ \emph {et~al.}(2012)\citenamefont
  {Abrikosov}, \citenamefont {Gorkov}, \citenamefont {Dzyaloshinski},\ and\
  \citenamefont {Silverman}}]{AbrikosovBook}%
  \BibitemOpen
  \bibfield  {author} {\bibinfo {author} {\bibfnamefont {A.}~\bibnamefont
  {Abrikosov}}, \bibinfo {author} {\bibfnamefont {L.}~\bibnamefont {Gorkov}},
  \bibinfo {author} {\bibfnamefont {I.}~\bibnamefont {Dzyaloshinski}}, \ and\
  \bibinfo {author} {\bibfnamefont {R.}~\bibnamefont {Silverman}},\ }\href@noop
  {} {\emph {\bibinfo {title} {Methods of Quantum Field Theory in Statistical
  Physics}}},\ Dover Books on Physics\ (\bibinfo  {publisher} {Dover
  Publications},\ \bibinfo {year} {2012})\BibitemShut {NoStop}%
\bibitem [{\citenamefont {Luttinger}(1960)}]{Luttinger:60}%
  \BibitemOpen
  \bibfield  {author} {\bibinfo {author} {\bibfnamefont {J.~M.}\ \bibnamefont
  {Luttinger}},\ }\href {\doibase 10.1103/PhysRev.119.1153} {\bibfield
  {journal} {\bibinfo  {journal} {Phys. Rev.}\ }\textbf {\bibinfo {volume}
  {119}},\ \bibinfo {pages} {1153} (\bibinfo {year} {1960})}\BibitemShut
  {NoStop}%
\bibitem [{\citenamefont {Hamann}\ and\ \citenamefont
  {Overhauser}(1966)}]{HamannOverhauser:66}%
  \BibitemOpen
  \bibfield  {author} {\bibinfo {author} {\bibfnamefont {D.~R.}\ \bibnamefont
  {Hamann}}\ and\ \bibinfo {author} {\bibfnamefont {A.~W.}\ \bibnamefont
  {Overhauser}},\ }\href {\doibase 10.1103/PhysRev.143.183} {\bibfield
  {journal} {\bibinfo  {journal} {Phys. Rev.}\ }\textbf {\bibinfo {volume}
  {143}},\ \bibinfo {pages} {183} (\bibinfo {year} {1966})}\BibitemShut
  {NoStop}%
\bibitem [{\citenamefont {Giuliani}\ and\ \citenamefont
  {Vignale}(2005{\natexlab{b}})}]{GiulianiVignaleFL:05}%
  \BibitemOpen
  \bibfield  {author} {\bibinfo {author} {\bibfnamefont {G.~F.}\ \bibnamefont
  {Giuliani}}\ and\ \bibinfo {author} {\bibfnamefont {G.}~\bibnamefont
  {Vignale}},\ }\enquote {\bibinfo {title} {The normal {F}ermi liquid},}\
  Chap.~\bibinfo {chapter} {8}, pp.\ \bibinfo {pages} {405--500},\ in\
  \cite{GiulianiVignale:05} (\bibinfo {year} {2005}{\natexlab{b}})\BibitemShut
  {NoStop}%
\bibitem [{\citenamefont {Brown}\ \emph
  {et~al.}(2013{\natexlab{a}})\citenamefont {Brown}, \citenamefont {Clark},
  \citenamefont {DuBois},\ and\ \citenamefont {Ceperley}}]{BrownCeperley:13a}%
  \BibitemOpen
  \bibfield  {author} {\bibinfo {author} {\bibfnamefont {E.~W.}\ \bibnamefont
  {Brown}}, \bibinfo {author} {\bibfnamefont {B.~K.}\ \bibnamefont {Clark}},
  \bibinfo {author} {\bibfnamefont {J.~L.}\ \bibnamefont {DuBois}}, \ and\
  \bibinfo {author} {\bibfnamefont {D.~M.}\ \bibnamefont {Ceperley}},\ }\href
  {\doibase 10.1103/PhysRevLett.110.146405} {\bibfield  {journal} {\bibinfo
  {journal} {Phys. Rev. Lett.}\ }\textbf {\bibinfo {volume} {110}},\ \bibinfo
  {pages} {146405} (\bibinfo {year} {2013}{\natexlab{a}})}\BibitemShut
  {NoStop}%
\bibitem [{\citenamefont {Brown}\ \emph
  {et~al.}(2013{\natexlab{b}})\citenamefont {Brown}, \citenamefont {DuBois},
  \citenamefont {Holzmann},\ and\ \citenamefont
  {Ceperley}}]{BrownCeperley:13b}%
  \BibitemOpen
  \bibfield  {author} {\bibinfo {author} {\bibfnamefont {E.~W.}\ \bibnamefont
  {Brown}}, \bibinfo {author} {\bibfnamefont {J.~L.}\ \bibnamefont {DuBois}},
  \bibinfo {author} {\bibfnamefont {M.}~\bibnamefont {Holzmann}}, \ and\
  \bibinfo {author} {\bibfnamefont {D.~M.}\ \bibnamefont {Ceperley}},\ }\href
  {\doibase 10.1103/PhysRevB.88.081102} {\bibfield  {journal} {\bibinfo
  {journal} {Phys. Rev. B}\ }\textbf {\bibinfo {volume} {88}},\ \bibinfo
  {pages} {081102} (\bibinfo {year} {2013}{\natexlab{b}})}\BibitemShut
  {NoStop}%
\bibitem [{\citenamefont {Brown}\ \emph
  {et~al.}(2013{\natexlab{c}})\citenamefont {Brown}, \citenamefont {DuBois},
  \citenamefont {Holzmann},\ and\ \citenamefont
  {Ceperley}}]{BrownCeperley:13berratum}%
  \BibitemOpen
  \bibfield  {author} {\bibinfo {author} {\bibfnamefont {E.~W.}\ \bibnamefont
  {Brown}}, \bibinfo {author} {\bibfnamefont {J.~L.}\ \bibnamefont {DuBois}},
  \bibinfo {author} {\bibfnamefont {M.}~\bibnamefont {Holzmann}}, \ and\
  \bibinfo {author} {\bibfnamefont {D.~M.}\ \bibnamefont {Ceperley}},\ }\href
  {\doibase 10.1103/PhysRevB.88.199901} {\bibfield  {journal} {\bibinfo
  {journal} {Phys. Rev. B}\ }\textbf {\bibinfo {volume} {88}},\ \bibinfo
  {pages} {199901} (\bibinfo {year} {2013}{\natexlab{c}})}\BibitemShut
  {NoStop}%
\bibitem [{\citenamefont {Schoof}\ \emph {et~al.}(2015)\citenamefont {Schoof},
  \citenamefont {Groth}, \citenamefont {Vorberger},\ and\ \citenamefont
  {Bonitz}}]{SchoofBonitz:15}%
  \BibitemOpen
  \bibfield  {author} {\bibinfo {author} {\bibfnamefont {T.}~\bibnamefont
  {Schoof}}, \bibinfo {author} {\bibfnamefont {S.}~\bibnamefont {Groth}},
  \bibinfo {author} {\bibfnamefont {J.}~\bibnamefont {Vorberger}}, \ and\
  \bibinfo {author} {\bibfnamefont {M.}~\bibnamefont {Bonitz}},\ }\href
  {\doibase 10.1103/PhysRevLett.115.130402} {\bibfield  {journal} {\bibinfo
  {journal} {Phys. Rev. Lett.}\ }\textbf {\bibinfo {volume} {115}},\ \bibinfo
  {pages} {130402} (\bibinfo {year} {2015})}\BibitemShut {NoStop}%
\bibitem [{\citenamefont {Malone}\ \emph {et~al.}(2016)\citenamefont {Malone},
  \citenamefont {Blunt}, \citenamefont {Brown}, \citenamefont {Lee},
  \citenamefont {Spencer}, \citenamefont {Foulkes},\ and\ \citenamefont
  {Shepherd}}]{MaloneShepherd:16}%
  \BibitemOpen
  \bibfield  {author} {\bibinfo {author} {\bibfnamefont {F.~D.}\ \bibnamefont
  {Malone}}, \bibinfo {author} {\bibfnamefont {N.~S.}\ \bibnamefont {Blunt}},
  \bibinfo {author} {\bibfnamefont {E.~W.}\ \bibnamefont {Brown}}, \bibinfo
  {author} {\bibfnamefont {D.~K.~K.}\ \bibnamefont {Lee}}, \bibinfo {author}
  {\bibfnamefont {J.~S.}\ \bibnamefont {Spencer}}, \bibinfo {author}
  {\bibfnamefont {W.~M.~C.}\ \bibnamefont {Foulkes}}, \ and\ \bibinfo {author}
  {\bibfnamefont {J.~J.}\ \bibnamefont {Shepherd}},\ }\href {\doibase
  10.1103/PhysRevLett.117.115701} {\bibfield  {journal} {\bibinfo  {journal}
  {Phys. Rev. Lett.}\ }\textbf {\bibinfo {volume} {117}},\ \bibinfo {pages}
  {115701} (\bibinfo {year} {2016})}\BibitemShut {NoStop}%
\bibitem [{\citenamefont {Dornheim}\ \emph {et~al.}(2016)\citenamefont
  {Dornheim}, \citenamefont {Groth}, \citenamefont {Sjostrom}, \citenamefont
  {Malone}, \citenamefont {Foulkes},\ and\ \citenamefont
  {Bonitz}}]{DornheimBonitz:16}%
  \BibitemOpen
  \bibfield  {author} {\bibinfo {author} {\bibfnamefont {T.}~\bibnamefont
  {Dornheim}}, \bibinfo {author} {\bibfnamefont {S.}~\bibnamefont {Groth}},
  \bibinfo {author} {\bibfnamefont {T.}~\bibnamefont {Sjostrom}}, \bibinfo
  {author} {\bibfnamefont {F.~D.}\ \bibnamefont {Malone}}, \bibinfo {author}
  {\bibfnamefont {W.~M.~C.}\ \bibnamefont {Foulkes}}, \ and\ \bibinfo {author}
  {\bibfnamefont {M.}~\bibnamefont {Bonitz}},\ }\href {\doibase
  10.1103/PhysRevLett.117.156403} {\bibfield  {journal} {\bibinfo  {journal}
  {Phys. Rev. Lett.}\ }\textbf {\bibinfo {volume} {117}},\ \bibinfo {pages}
  {156403} (\bibinfo {year} {2016})}\BibitemShut {NoStop}%
\bibitem [{\citenamefont {Karasiev}\ \emph {et~al.}(2014)\citenamefont
  {Karasiev}, \citenamefont {Sjostrom}, \citenamefont {Dufty},\ and\
  \citenamefont {Trickey}}]{KarasievTrickey:14}%
  \BibitemOpen
  \bibfield  {author} {\bibinfo {author} {\bibfnamefont {V.~V.}\ \bibnamefont
  {Karasiev}}, \bibinfo {author} {\bibfnamefont {T.}~\bibnamefont {Sjostrom}},
  \bibinfo {author} {\bibfnamefont {J.}~\bibnamefont {Dufty}}, \ and\ \bibinfo
  {author} {\bibfnamefont {S.~B.}\ \bibnamefont {Trickey}},\ }\href {\doibase
  10.1103/PhysRevLett.112.076403} {\bibfield  {journal} {\bibinfo  {journal}
  {Phys. Rev. Lett.}\ }\textbf {\bibinfo {volume} {112}},\ \bibinfo {pages}
  {076403} (\bibinfo {year} {2014})}\BibitemShut {NoStop}%
\bibitem [{\citenamefont {Groth}\ \emph {et~al.}(2017)\citenamefont {Groth},
  \citenamefont {Dornheim},\ and\ \citenamefont {Bonitz}}]{GrothBonitz:17}%
  \BibitemOpen
  \bibfield  {author} {\bibinfo {author} {\bibfnamefont {S.}~\bibnamefont
  {Groth}}, \bibinfo {author} {\bibfnamefont {T.}~\bibnamefont {Dornheim}}, \
  and\ \bibinfo {author} {\bibfnamefont {M.}~\bibnamefont {Bonitz}},\ }\href
  {\doibase 10.1002/ctpp.201600082} {\bibfield  {journal} {\bibinfo  {journal}
  {Contributions to Plasma Physics}\ }\textbf {\bibinfo {volume} {57}},\
  \bibinfo {pages} {137} (\bibinfo {year} {2017})}\BibitemShut {NoStop}%
\bibitem [{\citenamefont {Tanaka}\ and\ \citenamefont
  {Ichimaru}(1986)}]{TanakaIchimaru:86}%
  \BibitemOpen
  \bibfield  {author} {\bibinfo {author} {\bibfnamefont {S.}~\bibnamefont
  {Tanaka}}\ and\ \bibinfo {author} {\bibfnamefont {S.}~\bibnamefont
  {Ichimaru}},\ }\href {\doibase 10.1143/JPSJ.55.2278} {\bibfield  {journal}
  {\bibinfo  {journal} {Journal of the Physical Society of Japan}\ }\textbf
  {\bibinfo {volume} {55}},\ \bibinfo {pages} {2278} (\bibinfo {year}
  {1986})}\BibitemShut {NoStop}%
\bibitem [{\citenamefont {Singwi}\ \emph {et~al.}(1968)\citenamefont {Singwi},
  \citenamefont {Tosi}, \citenamefont {Land},\ and\ \citenamefont
  {Sj\"olander}}]{SingwiSjoelander:68}%
  \BibitemOpen
  \bibfield  {author} {\bibinfo {author} {\bibfnamefont {K.~S.}\ \bibnamefont
  {Singwi}}, \bibinfo {author} {\bibfnamefont {M.~P.}\ \bibnamefont {Tosi}},
  \bibinfo {author} {\bibfnamefont {R.~H.}\ \bibnamefont {Land}}, \ and\
  \bibinfo {author} {\bibfnamefont {A.}~\bibnamefont {Sj\"olander}},\ }\href
  {\doibase 10.1103/PhysRev.176.589} {\bibfield  {journal} {\bibinfo  {journal}
  {Phys. Rev.}\ }\textbf {\bibinfo {volume} {176}},\ \bibinfo {pages} {589}
  (\bibinfo {year} {1968})}\BibitemShut {NoStop}%
\bibitem [{\citenamefont {Spink}\ \emph {et~al.}(2013)\citenamefont {Spink},
  \citenamefont {Needs},\ and\ \citenamefont {Drummond}}]{SpinkDrummond:13}%
  \BibitemOpen
  \bibfield  {author} {\bibinfo {author} {\bibfnamefont {G.~G.}\ \bibnamefont
  {Spink}}, \bibinfo {author} {\bibfnamefont {R.~J.}\ \bibnamefont {Needs}}, \
  and\ \bibinfo {author} {\bibfnamefont {N.~D.}\ \bibnamefont {Drummond}},\
  }\href {\doibase 10.1103/PhysRevB.88.085121} {\bibfield  {journal} {\bibinfo
  {journal} {Phys. Rev. B}\ }\textbf {\bibinfo {volume} {88}},\ \bibinfo
  {pages} {085121} (\bibinfo {year} {2013})}\BibitemShut {NoStop}%
\bibitem [{\citenamefont {Holzmann}\ \emph {et~al.}(2003)\citenamefont
  {Holzmann}, \citenamefont {Ceperley}, \citenamefont {Pierleoni},\ and\
  \citenamefont {Esler}}]{HolzmannEsler:03}%
  \BibitemOpen
  \bibfield  {author} {\bibinfo {author} {\bibfnamefont {M.}~\bibnamefont
  {Holzmann}}, \bibinfo {author} {\bibfnamefont {D.~M.}\ \bibnamefont
  {Ceperley}}, \bibinfo {author} {\bibfnamefont {C.}~\bibnamefont {Pierleoni}},
  \ and\ \bibinfo {author} {\bibfnamefont {K.}~\bibnamefont {Esler}},\ }\href
  {\doibase 10.1103/PhysRevE.68.046707} {\bibfield  {journal} {\bibinfo
  {journal} {Phys. Rev. E}\ }\textbf {\bibinfo {volume} {68}},\ \bibinfo
  {pages} {046707} (\bibinfo {year} {2003})}\BibitemShut {NoStop}%
\bibitem [{\citenamefont {Perrot}\ and\ \citenamefont
  {Dharma-wardana}(1984)}]{PerrotDharma-wardana:84}%
  \BibitemOpen
  \bibfield  {author} {\bibinfo {author} {\bibfnamefont {F.}~\bibnamefont
  {Perrot}}\ and\ \bibinfo {author} {\bibfnamefont {M.~W.~C.}\ \bibnamefont
  {Dharma-wardana}},\ }\href {\doibase 10.1103/PhysRevA.30.2619} {\bibfield
  {journal} {\bibinfo  {journal} {Phys. Rev. A}\ }\textbf {\bibinfo {volume}
  {30}},\ \bibinfo {pages} {2619} (\bibinfo {year} {1984})}\BibitemShut
  {NoStop}%
\bibitem [{\citenamefont {Fetter}\ and\ \citenamefont
  {Walecka}(2003)}]{FetterWalecka}%
  \BibitemOpen
  \bibfield  {author} {\bibinfo {author} {\bibfnamefont {A.~L.}\ \bibnamefont
  {Fetter}}\ and\ \bibinfo {author} {\bibfnamefont {J.~D.}\ \bibnamefont
  {Walecka}},\ }\href@noop {} {\emph {\bibinfo {title} {Quantum Theory of
  Many-particle Systems}}},\ Dover Books on Physics\ (\bibinfo  {publisher}
  {Dover Publications},\ \bibinfo {year} {2003})\BibitemShut {NoStop}%
\bibitem [{\citenamefont {Hong}\ and\ \citenamefont
  {Mahan}(1994)}]{HongMahan:94}%
  \BibitemOpen
  \bibfield  {author} {\bibinfo {author} {\bibfnamefont {S.}~\bibnamefont
  {Hong}}\ and\ \bibinfo {author} {\bibfnamefont {G.~D.}\ \bibnamefont
  {Mahan}},\ }\href {\doibase 10.1103/PhysRevB.50.7284} {\bibfield  {journal}
  {\bibinfo  {journal} {Phys. Rev. B}\ }\textbf {\bibinfo {volume} {50}},\
  \bibinfo {pages} {7284} (\bibinfo {year} {1994})}\BibitemShut {NoStop}%
\bibitem [{\citenamefont {Bardeen}(1936)}]{Bardeen:36}%
  \BibitemOpen
  \bibfield  {author} {\bibinfo {author} {\bibfnamefont {J.}~\bibnamefont
  {Bardeen}},\ }\href {\doibase 10.1103/PhysRev.50.1093} {\bibfield  {journal}
  {\bibinfo  {journal} {Phys. Rev.}\ }\textbf {\bibinfo {volume} {50}},\
  \bibinfo {pages} {1098} (\bibinfo {year} {1936})}\BibitemShut {NoStop}%
\bibitem [{\citenamefont {Wigner}(1938)}]{Wigner:38}%
  \BibitemOpen
  \bibfield  {author} {\bibinfo {author} {\bibfnamefont {E.}~\bibnamefont
  {Wigner}},\ }\href {\doibase 10.1039/TF9383400678} {\bibfield  {journal}
  {\bibinfo  {journal} {Trans. Faraday Soc.}\ }\textbf {\bibinfo {volume}
  {34}},\ \bibinfo {pages} {678} (\bibinfo {year} {1938})}\BibitemShut
  {NoStop}%
\bibitem [{\citenamefont {Karasiev}\ \emph {et~al.}(2015)\citenamefont
  {Karasiev}, \citenamefont {Chakraborty},\ and\ \citenamefont
  {Trickey}}]{KarasievTrickey:15}%
  \BibitemOpen
  \bibfield  {author} {\bibinfo {author} {\bibfnamefont {V.~V.}\ \bibnamefont
  {Karasiev}}, \bibinfo {author} {\bibfnamefont {D.}~\bibnamefont
  {Chakraborty}}, \ and\ \bibinfo {author} {\bibfnamefont {S.}~\bibnamefont
  {Trickey}},\ }\href {\doibase http://doi.org/10.1016/j.cpc.2015.03.002}
  {\bibfield  {journal} {\bibinfo  {journal} {Computer Physics Communications}\
  }\textbf {\bibinfo {volume} {192}},\ \bibinfo {pages} {114 } (\bibinfo {year}
  {2015})}\BibitemShut {NoStop}%
\bibitem [{\citenamefont {Burke}\ \emph {et~al.}(2016)\citenamefont {Burke},
  \citenamefont {Smith}, \citenamefont {Grabowski},\ and\ \citenamefont
  {Pribram-Jones}}]{BurkePribramJones:16}%
  \BibitemOpen
  \bibfield  {author} {\bibinfo {author} {\bibfnamefont {K.}~\bibnamefont
  {Burke}}, \bibinfo {author} {\bibfnamefont {J.~C.}\ \bibnamefont {Smith}},
  \bibinfo {author} {\bibfnamefont {P.~E.}\ \bibnamefont {Grabowski}}, \ and\
  \bibinfo {author} {\bibfnamefont {A.}~\bibnamefont {Pribram-Jones}},\ }\href
  {\doibase 10.1103/PhysRevB.93.195132} {\bibfield  {journal} {\bibinfo
  {journal} {Phys. Rev. B}\ }\textbf {\bibinfo {volume} {93}},\ \bibinfo
  {pages} {195132} (\bibinfo {year} {2016})}\BibitemShut {NoStop}%
\bibitem [{Note1()}]{Note1}%
  \BibitemOpen
  \bibinfo {note} {The code used to obtain thermodynamic properties of the
  electron gas at finite temperature within the Hartree-Fock approximation can
  be downloaded from \protect \url
  {https://github.com/fgeich/pyFTEGhf}.}\BibitemShut {Stop}%
\bibitem [{\citenamefont {Eich}\ \emph {et~al.}(2014)\citenamefont {Eich},
  \citenamefont {Di~Ventra},\ and\ \citenamefont {Vignale}}]{EichVignale:14a}%
  \BibitemOpen
  \bibfield  {author} {\bibinfo {author} {\bibfnamefont {F.~G.}\ \bibnamefont
  {Eich}}, \bibinfo {author} {\bibfnamefont {M.}~\bibnamefont {Di~Ventra}}, \
  and\ \bibinfo {author} {\bibfnamefont {G.}~\bibnamefont {Vignale}},\ }\href
  {\doibase 10.1103/PhysRevLett.112.196401} {\bibfield  {journal} {\bibinfo
  {journal} {Phys. Rev. Lett.}\ }\textbf {\bibinfo {volume} {112}},\ \bibinfo
  {pages} {196401} (\bibinfo {year} {2014})}\BibitemShut {NoStop}%
\bibitem [{\citenamefont {Eich}\ \emph {et~al.}(2017)\citenamefont {Eich},
  \citenamefont {Ventra},\ and\ \citenamefont {Vignale}}]{EichVignale:17}%
  \BibitemOpen
  \bibfield  {author} {\bibinfo {author} {\bibfnamefont {F.~G.}\ \bibnamefont
  {Eich}}, \bibinfo {author} {\bibfnamefont {M.~D.}\ \bibnamefont {Ventra}}, \
  and\ \bibinfo {author} {\bibfnamefont {G.}~\bibnamefont {Vignale}},\ }\href
  {http://stacks.iop.org/0953-8984/29/i=6/a=063001} {\bibfield  {journal}
  {\bibinfo  {journal} {Journal of Physics: Condensed Matter}\ }\textbf
  {\bibinfo {volume} {29}},\ \bibinfo {pages} {063001} (\bibinfo {year}
  {2017})}\BibitemShut {NoStop}%
\end{thebibliography}%

\end{document}